\documentclass[
twocolumn,
]{ceurart}
\makeatletter
\DeclareOldFontCommand{\rm}{\normalfont\rmfamily}{\mathrm}
\DeclareOldFontCommand{\sf}{\normalfont\sffamily}{\mathsf}
\DeclareOldFontCommand{\tt}{\normalfont\ttfamily}{\mathtt}
\DeclareOldFontCommand{\bf}{\normalfont\bfseries}{\mathbf}
\DeclareOldFontCommand{\it}{\normalfont\itshape}{\mathit}
\DeclareOldFontCommand{\sl}{\normalfont\slshape}{\@nomath\sl}
\DeclareOldFontCommand{\sc}{\normalfont\scshape}{\@nomath\sc}
\makeatother

\usepackage{float}
\usepackage{booktabs}
\usepackage{tabularx}
\usepackage[english]{babel}
\usepackage[capitalize, noabbrev]{cleveref}

\usepackage{graphicx, caption, subcaption}
\usepackage{todonotes}
\usepackage[export]{adjustbox}

\usepackage{url}

\usepackage{breakurl}
\PassOptionsToPackage{breaklinks}{hyperref}

\newcommand{\lastaccessed}{\textit{last accessed 17.12.2020}}

\definecolor{DanielsColor}{rgb}{0.9,0.6,0.1}

\definecolor{HaisColor}{rgb}{0.9,0.3,0.9}

\definecolor{FloriansColor}{rgb}{0,0.3,0.9}

\definecolor{LukasColor}{rgb}{0.3,0.9,0.3}

\begin{document}

\copyrightyear{2021}
\copyrightclause{Copyright for this paper by its authors. Use permitted under Creative Commons License Attribution 4.0 International (CC BY 4.0).}

\conference{Joint Proceedings of the ACM IUI 2021 Workshops, April 13-17, 2021, College Station, USA}

\title{Nine Potential Pitfalls when Designing Human-AI Co-Creative Systems}

\author[1]{Daniel Buschek}
\ead{daniel.buschek@uni-bayreuth.de}

\author[2,3]{Lukas Mecke}
\ead{lukas.meckek@unibw.de}

\author[1]{Florian Lehmann}
\ead{florian.lehmann@uni-bayreuth.de}

\author[1]{Hai Dang}
\ead{hai.dang@uni-bayreuth.de}

\address[1]{Research Group HCI + AI, Department of Computer Science, University of Bayreuth, Bayreuth, Germany
}
\address[2]{Bundeswehr University Munich, Munich, Germany
}
\address[3]{LMU Munich, Munich, Germany
}

\begin{abstract}
This position paper examines \textit{potential pitfalls} on the way towards achieving human-AI co-creation with generative models in a way that is beneficial to the users' interests.
In particular, we collected a set of nine potential pitfalls, based on the literature and our own experiences as researchers working at the intersection of HCI and AI. We illustrate each pitfall with examples and suggest ideas for addressing it. 
Reflecting on all pitfalls, we discuss and conclude with implications for future research directions.
With this collection, we hope to contribute to a critical and constructive discussion on the roles of humans and AI in co-creative interactions, with an eye on related assumptions and potential side-effects for creative practices and beyond. 
\end{abstract}


\maketitle

\section{Introduction}

Ongoing advances in generative AI systems have sparked great interest in using them interactively in creative contexts and for digital content creation and manipulation:
Some examples include (1) generating or modifying images with generative adversarial networks (GANs)~\cite{Bau2020,Harkonen2020,Karras2020}, (2) generating controllable movements for virtual characters with recurrent neural networks, deep reinforcement learning and physics simulations~\cite{Park2019}, and (3) controllable machine capabilities for generating or summarizing when working with text~\cite{Dathathri2020,Gehrmann2020}.
Such computational methods have also entered specifically artistic domains, including visual art~\cite{Akten2019}, creative writing and poetry~\cite{Gero2019, Ghazvininejad2017}. More examples can be found in a curated ``ML x Art'' list\footnote{\url{https://mlart.co/}, \lastaccessed}.

A common vision, also present in the call for this workshop, paints a picture of creative human use of such AI as tools. In this view, these new interactive systems are hoped to realise key ideas from \textit{creativity support tools} (CST,~\cite{Frich2019}) by leveraging AI capabilities. 
More specifically, this support could cast humans and AI in many different roles (for a recent overview see \cite{Kantosalo2020roleoverview}). This includes, for example, using AI as a divergent or convergent agent, as described by \citet{Hoffmann2016}, that is, to generate or evaluate (human) ideas. Related, \citet{Kantosalo2016} highlight alternating co-creation, with the AI ``pleasing'' and ``provoking'' the user. Moreover, \citet{NegreteYankelevich2014} describe a related set of roles, including AI as an ``apprentice'', whose work is judged and selectively chosen by humans, or a leader-like role, which only leaves final configurations to the user.

Within this range of roles, the workshop call emphasises the \textit{generative} capabilities of AI. In this paper, we thus focus on the role of AI as a generator, and the underlying goal of freeing its users to focus on a larger creative vision, while the AI takes care of more tedious steps.

With this goal in mind, this paper examines \textit{potential pitfalls} on the way towards achieving it in practice. 
Our research approach is related to work on \textit{dark patterns} in UI/UX design~\cite{Gray2018}, which also examines -- sometimes speculatively~\cite{Chromik2019}, sometimes empirically~\cite{DiGeronimo2020} -- what ``could go wrong'', in order to ultimately inspire directions for interaction design that are beneficial to the users' interests. 
In doing so, we thus hope to contribute to a critical and constructive discussion on the roles of humans and AI in co-creative interactions, with an eye on related assumptions and potential side-effects for creative practices and beyond.

\section{Research Approach}

Our interest in collecting pitfalls is inspired by work on dark patterns~\cite{Chromik2019, DiGeronimo2020, Gray2018}: Both pitfalls and dark patterns identify issues with user interfaces and interactions that result in experiences or outcomes which might not be in the user's best interest. However, in contrast to what is often assumed in dark patterns, pitfalls do not imply bad intention, rather oversight or lack of information\footnote{\url{https://www.merriam-webster.com/dictionary/pitfall}, \lastaccessed}.

Concretely, related work collected speculative dark patterns for explainability, transparency and control in intelligent interactive systems~\cite{Chromik2019} by transferring dark patterns previously described for UI/UX design~\cite{Gray2018}. Other work collected dark UI/UX patterns empirically by reviewing a large set of existing mobile applications~\cite{DiGeronimo2020}.
Both approaches seem challenging to directly transfer to collecting pitfalls in the context of co-creative generative AI, since there are no previously defined pitfalls and no easily accessible collections (or ``app stores'') of many usable applications for review.

Therefore, we followed a qualitative, speculative approach and brainstormed on potential pitfalls, or \textit{``what could go wrong''} (cf.~\cite{Colusso2019}) in interactions with co-creative AI. Here we are loosely inspired by aspects of speculative design~\cite{Dunne2013}, although that area typically aims to address broader issues than what we focus on here. Further inspiring ``speculative futures'' for human-AI co-creative systems, along with a conceptual framework, can be found in the work by \citet{Bown2018}.

We particularly explore issues grounded in today's interactions and UIs, which can be reasonably well imagined to potentially occur with the current state of the art of generative AI models.
In particular, our brainstorming started from three prompts: (1) Issues arising from currently \textit{limited capabilities of AI}, and (2) from exploring what might happen with \textit{too much AI involvement}; plus (3) thinking \textit{beyond use} and usage situations.
Considering this approach, we see the pitfalls presented here not as a comprehensive and ``definitive'' list but rather as a stimulus for discussion in the research community -- at the workshop and beyond.

\begin{table*}
\centering
\scriptsize
\newcolumntype{P}[1]{>{\raggedright\let\newline\\\arraybackslash\hspace{0pt}}p{#1}}
\newcolumntype{L}{>{\raggedright\let\newline\\\arraybackslash\hspace{0pt}}X}
\renewcommand{\arraystretch}{1.75}
\setlength{\tabcolsep}{2pt}
\begin{adjustbox}{center}
\begin{tabularx}{1.2\textwidth}{@{}P{.5em}P{6em}P{6em}LLLL@{}}
\toprule
& \footnotesize\textbf{Name} & \footnotesize\textbf{Affected aspects} & \footnotesize\textbf{Problem description} & \footnotesize\textbf{Example} & \footnotesize\textbf{How it might have happened} (examples) & \footnotesize\textbf{How it might be addressed} (examples)\\ 
\midrule

\multicolumn{7}{c}{\small\textbf{Limited AI}}\\

1 & Invisible AI boundaries & model, creativity, exploration & A (generative) AI component imposes unknown restrictions on creativity and exploration. & An AI face image editor cannot make faces bald without also turning them male-looking. & Model with limited generalisability beyond training data, and entangled or nonsensical (latent) dimensions w.r.t. human understanding. & UI: Show boundaries e.g. via uncertainty, samples, precision/recall~\cite{Kynkaanniemi2019}. AI: Improve generalisabilty, disentanglement; consider narrowing scope.\\

2 & Lack of expressive interaction & usability, creativity, exploration & The UI imposes a ``bottleneck'' on creative use of the AI. & Image generator is controlled with many 1D inputs for a high D latent space~\cite{harkonen2020ganspace} - vs. rich image editor tools like brushes. & Fine-grained AI control is difficult. ``Conservative'' UI design focused on ensuring input stays in (training) data distribution. & Human-centred design with target group, e.g. to inform preferable tradeoffs of UI expressiveness and model ``breaking points''.\\

3 & False sense of proficiency & trust, reliability & AI suggests answers or completions that the user cannot verify or that generate a false sense of proficiency. & When prompted to complete a sentence about the population of a large city the AI delivers a reasonable number that could be correct -- but might not be. & Language models are capable of memorizing excerpts of text and reproducing them when prompted with a similar context. %
& Learn an additional model, that can attribute generated content to an explicit source to allow for verifying correctness.\\

\midrule
\multicolumn{7}{c}{\small\textbf{Too much AI}}\\

4 & Conflicts of territory & usability, UX, control & AI overwrites what the user has manually created/edited. & In a co-creative text editor, the user replaces terms in generated text. Later, the AI (partly) reverts these changes. & Language model optimised for word probability and user's term was less likely. & Keep track of user edits to protect them, ask for confirmation before changes, or to integrate this info into inference. \\

5 & Agony of choice & usability, UX, productivity & AI provides over\-whelming amount/detail of content that distracts or creates agony of choice. & An AI photo editor displays an excessive number of suggested variants. The resulting small previews make it hard to discern and decide. & UI design process was focused on showing AI capabilities instead of user needs. & Clarifying use cases and support needs, re\-spon\-sive / malleable UI concepts, changeable user settings.\\ 

6 & Time waster & usability, UX, productivity & AI interrupts user or draws attention away from the creative task itself. & A co-creative music composition tool continuously shows melody completions, which keep the user busy with exploring or understanding the system instead of their ideas. & Same as above. Also: Timing of the AI's involvement not tested with users or varying preferences between users. & Same as above. Attention-aware UI (e.g. AI waits to not disrupt user's focused work; or stops suggestions if user has explored it for a while). \\ 

\midrule
\multicolumn{7}{c}{\small\textbf{Beyond use}}\\

7 & AI bias & accountability, fairness, transparency & AI suggestions are biased in a certain unwanted way, w.r.t. human meaning and values. & An AI story generator writes gender-stereotypical protagonists (e.g. w.r.t. roles/occupations). & AI picked up biases in the training data or created bias through its learning method. Development process unaware of biases. & Design for easy human revision/rejection. Addressing AI bias (e.g. see \cite{pmlr-v81-buolamwini18a, shah-etal-2020-predictive}). Learning from user feedback/actions. \\

8 & Conflict of Creation \& Responsibility & creativity, responsibility, ownership & A system and a user collaborate to create an output. Ownership and responsibility are unclear. & In a co-creative text editor the AI suggests formulations that appear verbatim in the training data. Who is the owner of the resulting text? & Co-creative systems operate on a continuum between user and system creation%
, challenging attributions of ownership. & Should we attribute an AI and training data providers as contributors? Do we need systems to check for (accidental) plagiarism?\\ 

9 & User and Data Privacy & privacy, responsibility  & Private data may be exposed through the AI system or its training data. & 1) A user \textit{A} works with a cloud-based AI text creator and their data is transmitted unencrypted. 2) The AI reveals (private parts of) another user \textit{B's} data to \textit{A} (e.g.~\cite{carlini2018secret, carlini2020extracting}). & AI models are trained on a large corpus of data and can sometimes default to replicating this data when prompted. & Remove private information from training sets and work with AI either encrypted or locally. \\ 

\bottomrule
\end{tabularx}
\end{adjustbox}
\caption{Overview of the collected pitfalls. Additionally, Figure~\ref{fig:examples} visualises one example for each of the categories ``Limited AI'', ``Too much AI'' and ``Beyond use''.}
\label{tab:overview}
\end{table*}

\section{Nine Potential Pitfalls}

\cref{tab:overview} shows the pitfalls we collected. In particular, we present nine pitfalls, three for each of our starting prompts, that is, for limited AI (pitfalls 1-3), too much AI involvement (pitfalls 4-6), and for aspects beyond use (pitfalls 7-9).

\begin{figure*}
    \centering
    \begin{subfigure}[b]{0.43\linewidth}
        \includegraphics[width=\linewidth,trim=0 -3 0 -3, clip,cfbox=lightgray 0.5pt 0.5pt]{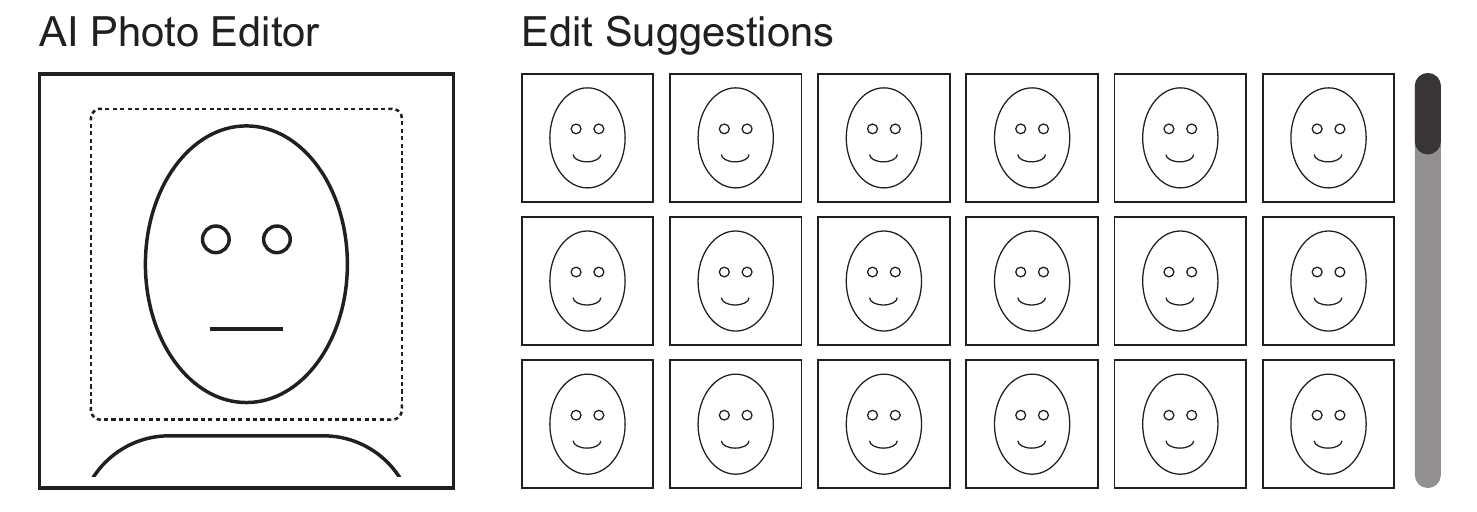}
        \subcaption{An AI photo editor displays an excessive number of suggestions. Due to the number of options and the small previews (making it hard to see what each option achieves) the user is left in an agony of choice. }
    \end{subfigure}
    \hspace{10pt}
    \begin{subfigure}[b]{0.43\linewidth}
        \includegraphics[width=\linewidth,trim=70 0 69 85, clip,cfbox=lightgray 0.5pt 0.5pt]{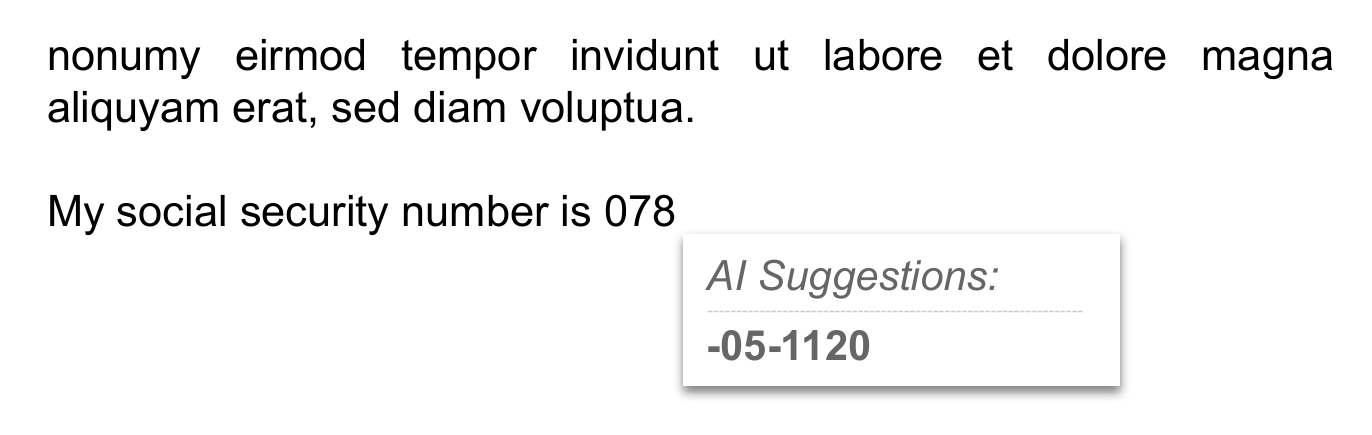}
        \subcaption{Example for an AI leaking sensitive information from the training dataset (based on \cite{carlini2018secret}), either as a suggestion or as a response to a primer (enabling active attacks). Such an attack has been demonstrated by~\citet{carlini2020extracting}. }
    \end{subfigure}
    \vspace{10pt}
    \begin{subfigure}[b]{\linewidth}
        \centering
        \includegraphics[width=0.75\linewidth,trim=30 10 30 10,clip,cfbox=lightgray 0.5pt 0.5pt]{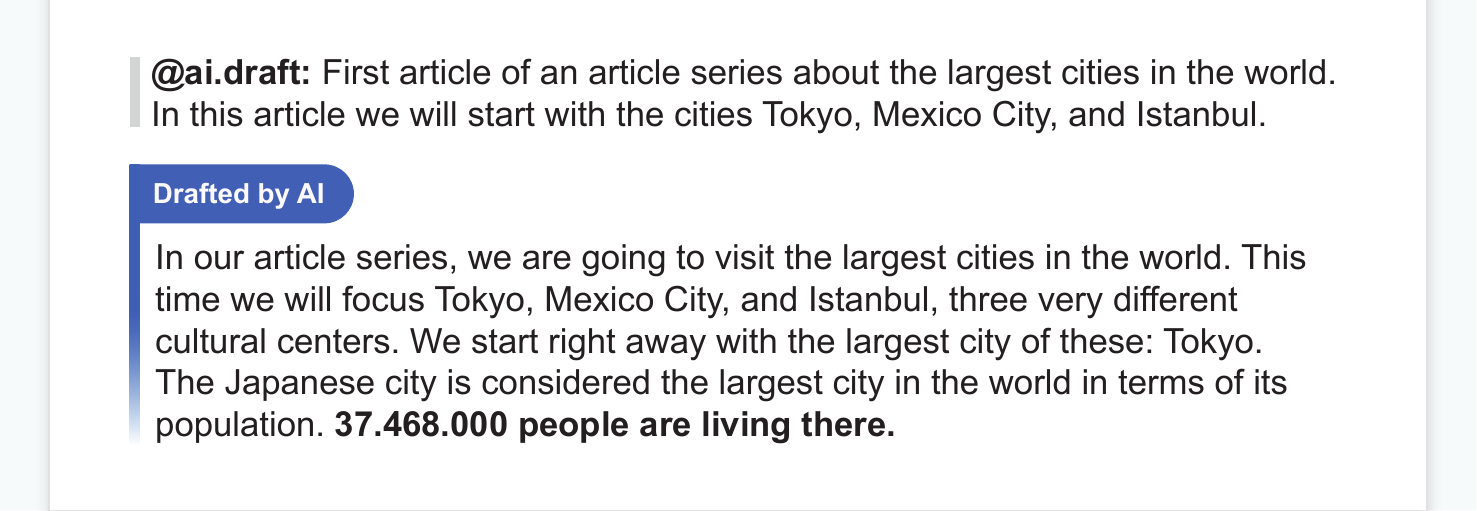}
        \subcaption{A text editing tool could offer intelligent features, e.g. drafting paragraphs or completing a sentence. Yet, the AI might not have the capability to refer to sources -- to the human it remains unclear if the claims in a text are true. This leads to a false sense of proficiency. Here, the AI drafted a paragraph with claims about Tokyo's approximate population (bold). However, it refers to the metropolitan area, not the city proper. The interface in the figure is inspired by \citet{Yang2019}.}
    \end{subfigure}
    
    \caption{Collection of visual examples for the pitfalls shown in Table \ref{tab:overview}. Here we show potential interfaces and situations in which selected pitfalls may occur, leading to (a) agony of choice, (b) a breach of privacy or (c) a false sense of proficiency.}
    \label{fig:examples}
\end{figure*}

The table characterises each pitfall with a name, affected aspects (categories), a description of the problem, and a concise pitfall ``vignette'': This includes an example scenario describing a system in which this issue arises, along with an illustrating diagnosis of how this might have happened in the design and development of said system, plus corresponding ideas for potential solutions or open questions. For each category of pitfalls (``limited AI'', ``too much AI'', and ``beyond use'') we picked one example for further illustration in Figure~\ref{fig:examples}.

\begin{figure*}
    \centering
    \includegraphics[width=\linewidth]{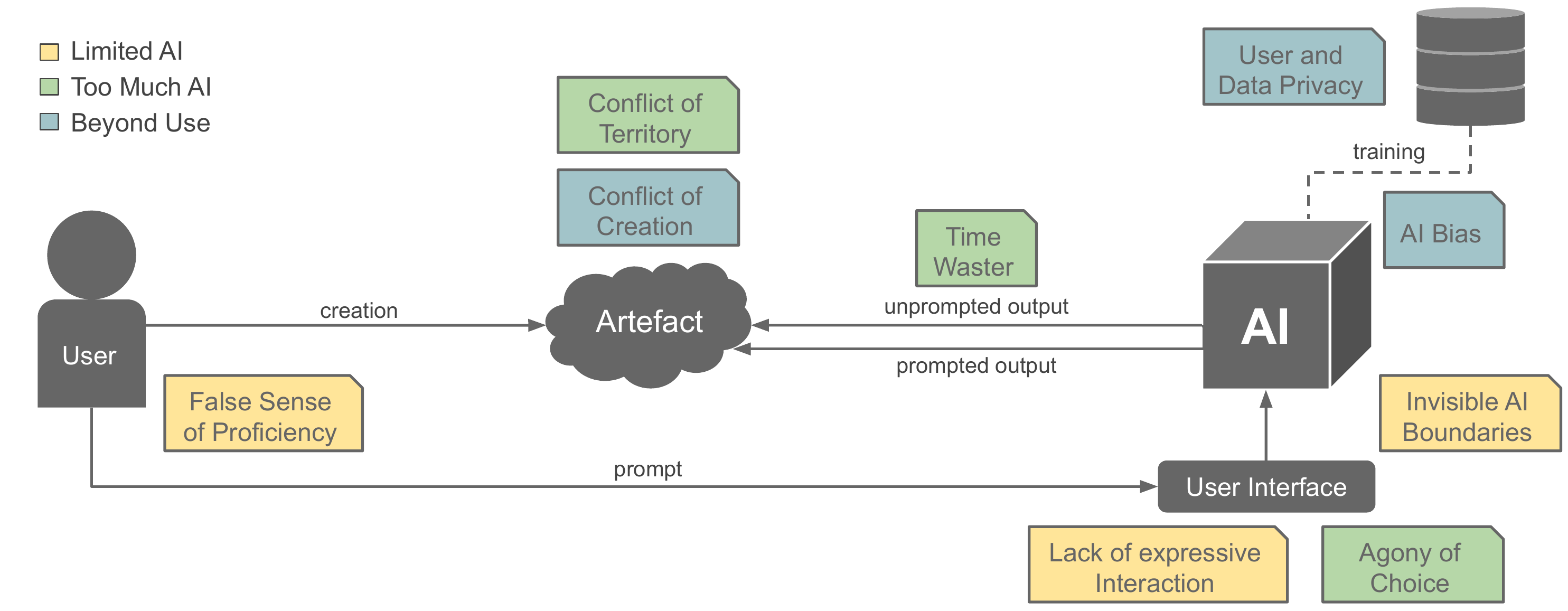}
    \caption{Visualisation of our underlying mental model of the interaction loop in human-AI co-creative systems. We place our identified pitfalls (see Table \ref{tab:overview}) in this loop based on the position where they most likely occur.}
    \label{fig:interaction_loop}
\end{figure*}

As an additional overview, \cref{fig:interaction_loop} locates these pitfalls within an interaction loop in human-AI co-creative systems; the loop is taken from a framework by \citet{Guzdial2019}. 
In this figure we illustrate our underlying mental model of human-AI interaction. It consists of the \textit{user} and the \textit{AI} as potential actors collaborating on a shared \textit{artifact}. The AI can get involved in the creation process in one of two ways: It can either be prompted to contribute through the user interface (e.g. using a predefined function to achieve an image manipulation) or it can act without a (user) prompt, e.g. to suggest edits or flag errors. We further include the training data in this model, as it provides the basis for the AI's actions and decisions.  
While we located the pitfalls within this model, these locations are by no means the only possible ones. They represent our interpretations of which point in the interaction loop is most likely affected by each pitfall. As an example, a lack of expressive interaction may not only be rooted in the user interface, but can also be caused by insufficient training data to support more meaningful options.

\section{Discussion}

\subsection{What are the Consequences of these Pitfalls?}

While \cref{tab:overview} lists concrete example problems, here we reflect more broadly on the consequences of such pitfalls for co-creative generative systems. In particular, we see two broad directions -- overt and covert consequences.

First, users might be annoyed, distracted, or otherwise put off by bad user experiences due to these pitfalls. For example, cases where the AI directly overwrites the user (pitfall 4), or distracts the user from their productive task (pitfall 6) might be particularly harmful in this regard. Observing AI failures might lead to \textit{algorithm aversion}, as described by \citet{Dietvorst2015}. In these cases, users might avoid future use of such systems.

In contrast, users might also be affected negatively \textit{without} noticing it. For example, this might be the case if the AI implies invisible boundaries (pitfall 1) that hinder creative exploration. Similarly, ``silent'' issues might result from the generative AI introducing incorrect information (pitfall 3), distractions (pitfall 6), or biases and legal issues (pitfalls 7-9). Users might only (much later) stumble across issues in downstream processes, evaluations or reflections. If such issues then affect evaluation of the user's creative work (e.g. due to false information, pitfall 3), this might result in \textit{algorithm anxiety}, described by~\citet{Jhaver2018}. 

Overall, the pitfalls might thus result in a range of possible consequences, from bad user experiences, negative impacts on creative work, abandonment of tools, to broader issues, including privacy related and legal ones.

\subsection{How can the Pitfalls Inform Research and Design of Co-Creative Generative Systems?}

Put briefly, this position paper describes what could go wrong in order to stimulate discussions of how to get it right. More concretely, here we describe three potential uses.

\subsubsection{Raising Awareness of Design Considerations}
The described pitfalls can help researchers and designers to think about a wide range of concrete aspects of interaction and UI design for co-creative generative systems (e.g. temporal and spatial integration of AI actions in UIs). In this way, they may \textit{raise awareness for making design choices explicit} that might have otherwise not been prominently considered. These design choices could then also be considered in light of relevant frameworks, such as Horvitz' mixed initiative principles~\cite{Horvitz1999} or the co-creative framework described by \citet{Guzdial2019} (cf. Figure~\ref{fig:interaction_loop}).

\subsubsection{Informing Comparisons and Baselines}
Moreover, the problematic systems described in the pitfalls in \cref{tab:overview} might \textit{inspire informative baseline systems for comparison} with (hopefully) better solutions. For example, a typical HCI user study on an AI photo editor might compare an AI vs non-AI version. However, as illustrated with the example for pitfall 5 (Agony of Choice), another insightful evaluation might further use a baseline that involves AI ``even more'' than the intended design solution to be evaluated.

\subsubsection{Making the Criteria for Successful Design Explicit}
Evaluating technical systems for creative use is challenging~\cite{Lamb2018}, for example, since creativity and quality criteria are often hard to operationalise, and may require interdisciplinary knowledge. Additionally involving AI can be expected to complicate evaluations further. Here, our pitfalls and examples may provide helpful concrete starting points, as a \textit{thinking prompt towards developing evaluations} that satisfy both HCI and AI interests. 
For instance, readers and workshop participants (with different backgrounds) could think about how they would improve the design -- and evaluate it -- for a concrete problematic example system in \cref{tab:overview}; and in particular how they might then make explicit and formulate their criteria in these cases.

\subsection{Will the Pitfalls Vanish with Better AI?}
One may ask if the illustrated issues might simply vanish in future systems that can build on better AI capabilities. Based on our considerations here, we do not expect this to be case: Co-creative systems involving both human and AI actions are not only limited by AI capabilities. We also have to expect problems arising from interaction and UI design as well as from integration into creative human practices. For example, a lack of expressiveness in interactions (pitfall 2) can still cause problems for creative human use, even in a system with a powerful, ``perfect'' generative model under the hood. 

In summary, the pitfalls highlight that human-AI co-creative systems sit at the intersection of HCI and AI, and that successful designs need to consider human-centred aspects in the process.
Our pitfalls reflect this in their mix of issues relating to interaction, UI and AI. We thus aim to motivate interdisciplinary work on such systems, also regarding research and design methodology.

\section{Conclusion}

One vision of interactive use of AI tools in co-creative settings focuses on the role of the AI as a generator that augments what people can achieve in creative tasks. This paper examined \textit{potential pitfalls} on the way towards achieving this vision in practice, starting from three speculation prompts: Issues arising from  (1) limited AI, (2) too much AI involvement, and (3) thinking beyond use and usage situations.

Concretely, we collected a set of nine potential pitfalls (\cref{tab:overview}) and discussed possible consequences and takeaways for researchers and designers along with illustrating examples. 
With this collection, we hope to contribute to a critical and constructive discussion on the roles of humans and AI in co-creative interactions, with an eye on related assumptions and potential side-effects for creative practices and beyond.

\begin{acknowledgments}
This project is funded by the Bavarian State Ministry of Science and the Arts and coordinated by the Bavarian Research Institute for Digital Transformation (bidt).
\end{acknowledgments}

\balance
\bibliography{bibliography}

\end{document}